\documentclass{iau} 
\usepackage{graphicx}

\title[Dwarf Galaxies as Cosmological Probes] 
{Dwarf Galaxies as Cosmological Probes}

\author[Julio F. Navarro]   
{Julio F. Navarro$^1$
}

\affiliation{$^1$CIfAR Senior Fellow and Professor. Department of Physics and Astronomy, University of
  Victoria, Victoria, BC, Canada V8P 5C2 \\ email: {\tt jfn@uvic.ca} \\[\affilskip]
}

\pubyear{2018}
\volume{344}  
\jname{Dwarf Galaxies: from the Deep Universe to the Present}
\editors{Kristen McQuinn \& Sabrina Stierwalt, eds.}
\begin{document}

\maketitle

\begin{abstract}
  The Lambda Cold Dark Matter (LCDM) paradigm makes specific
  predictions for the abundance, structure, substructure and
  clustering of dark matter halos, the sites of galaxy
  formation. These predictions can be directly tested, in the low-mass
  halo regime, by dark matter-dominated dwarf galaxies. A number of
  potential challenges to LCDM have been identified when confronting
  the expected properties of dwarfs with observation. I review our
  understanding of a few of these issues, including the ``missing
  satellites'' and the ``too-big-to-fail'' problems, and argue that
  neither poses an insurmountable challenge to LCDM. Solving these
  problems requires that most dwarf galaxies inhabit halos of similar
  mass, and that there is a relatively sharp minimum halo mass threshold to
  form luminous galaxies. These predictions are eminently
  falsifiable. In particular, LCDM predicts a large number of ``dark''
  low-mass halos, some of which should have retained enough primordial
  gas to be detectable in deep 21 cm or H$_\alpha$ surveys. Detecting
  this predicted population of ``mini-halos'' would be a major
  discovery and a resounding success for LCDM on small scales.
  \keywords{Cosmology: dark matter, Galaxies: dwarf, Galaxies:
    formation.}
\end{abstract}

\firstsection 
\section{Introduction}
\label{SecIntro}

There is now a well-defined paradigm for the growth of structure in
the Universe. Large-scale observations of the cosmic microwave
background and of galaxy clustering have helped to constrain the
matter-energy content of the Universe, as well as its expansion
history and overall geometry. These observations are well reproduced
in the LCDM paradigm, which assumes that the recent accelerated
expansion of the Universe is due to the ``dark energy'' contribution
of a cosmological constant (``Lambda'', or ``L''); that the matter
content of the Universe is dominated by cold dark matter (CDM), and
that the initial density fluctuation was scale-free and Gaussian, as
expected from inflation \cite[(Planck Collaboration
2016)]{Planck2016}.

With the cosmological parameters settled, it is then possible to
predict and/or simulate the present-day clustering of dark matter on
all extragalactic scales relevant to the formation of galaxies, a
field where large cosmological N-body simulations have become
indispensible. On very large scales, dark matter distributes itself in
a foam-like network of sheets and filaments. This ``cosmic web'' is
punctuated by high-density regions (``halos'') where the dynamical
crossing time is shorter than the age of the universe. Thanks largely
to N-body simulations, we now have an excellent understanding of the
expected structure, abundance, and clustering of cold dark matter
halos in LCDM (see; e.g., \cite[Frenk \& White 2012]{FrenkWhite2012} for a recent review).

CDM halos are expected to be self-similar in structure and
substructure. In terms of structure, the mass profile of all halos,
regardless of mass, look alike when scaled properly (\cite[Navarro et
al 1996, 1997]{NFW96,NFW97}). In terms of substructure, the similarity
implies that the mass function of subhalos (``satellites'') is
independent of halo mass when scaled to the mass of the host (see;
e.g., \cite[Moore et al 1999, Wang et al
2012]{Moore1999,Wang2012}). 

These are important results. If, when scaled, all halos look alike,
then, unscaled, all halos should look different.  For example, the
density profiles of two LCDM halos of different mass do not cross at
any radius. This implies that measuring the halo properties at {\it one}
radius (e.g., density, circular velocity, etc) allows a full
characterization of the halo at {\it all} radii. In terms of substructure,
it implies that the expected numbers of subhalos of a given system
scales directly with the system's total mass. In other words, a more
massive halo has more subhalos of all masses.

In the LCDM paradigm, galaxies are assumed to form at the centers of
dark matter halos and satellites are assumed to inhabit their
subhalos. Since the mass function of CDM halos and the stellar mass
function of galaxies are both known, the above assumption implies that
it should be possible to infer the relation between the stellar mass
of a galaxy and the mass of its surrounding halo using simple, but
robust, approximations. The most widely used approximation is referred
to as ``abundance-matching'' (AbMat), where galaxies are assigned to
halos respecting their relative ranks, by mass (see, e.g.,
\cite[Behroozi et al 2013]{Behroozi2013}, and references
therein). Thus, in a given (large!) volume, the most massive galaxy
should inhabit the most massive halo, the 2nd most massive galaxy the
2nd most massive halo, and so on. Because the {\it shape} of the
galaxy and halo mass functions are quite different, this assignment
results in a strongly non-linear dependence of galaxy mass on halo
mass.

In particular, because the faint end of the luminosity function is
relatively flat, and the low-mass end of the halo mass function is
rather steep, this implies that {\it most} dwarf galaxies are assigned
to halos of similar mass. Indeed, the AbMat model predicts that dwarfs
spanning a factor of 10,000 in stellar mass
($10^5<M_{\rm gal}/M_\odot<10^9$) all inhabit halos spanning a rather
narrow range of halo virial\footnote{We use a mean density of
  $200\times$ the critical density to define the virial boundary of a
  halo.} mass, roughly between
$10^{10}<M_{200}/M_\odot<10^{11}$. Further, it predicts that few {\it
  field}\footnote{Satellites may lose large fractions of their dark
  matter to tidal stripping, so this statement does not apply to the
  halo masses of satellites of more massive systems.} luminous
galaxies must form in halos less massive than $10^{10}\, M_\odot$, and
essentially {\it none} in halos under $10^9 \, M_\odot$.

The latest cosmological hydrodynamical simulations of galaxy formation
in LCDM have confirmed the AbMat predictions (see; e.g.,
\cite[Vogelsberger et al 2014, Schaye et al 2015]{Vogelsberger2014,Schaye2015}), at least in the $M_{\rm
  gal}>10^{7}$-$10^{8}\, M_\odot$ regime, where some of the simulations are able
to reproduce the galaxy stellar mass function to within a factor of
two. It is unclear whether this success extends to lower masses, since
the galaxy mass function is not well constrained on that regime and
the simulations have insufficient resolution to provide robust results
on cosmologically significant volumes. 

At the very faint end, the shape of the luminosity function is
probably best constrained in the Local Group (LG); the loose
association of the Milky Way (MW) and Andromeda (M31) galaxies, their
satellites, and surrounding field. The inventory of LG galaxies
brighter than the Draco dwarf spheroidal (dSph), which has an absolute
magnitude of $M_V\sim -8$ and a stellar mass
$M_{\rm gal}\sim 10^5\, M_\odot$, is thought to be quite complete, at
least within $2$-$3$ Mpc from the MW-M31 barycenter (\cite[McConnachie
2012]{McConnachie2012}). It is also possible to measure the kinematics
of these nearby dwarfs, enabling meaningful constraints on their dark
matter content (and hence on their halo masses), at least in the
region where kinematic tracers exist.

These facts make the Local Group an ideal environment to test the
relation between galaxy mass and halo mass predicted by LCDM, and have
prompted a number of groups to simulate volumes tailored to reproduce
either the overall configuration of the massive galaxies of the Local
Group (see; e.g., the ELVIS and APOSTLE projects;
\cite[Garrison-Kimmel et al. 2014, Fattahi et
al. 2016a]{Garrison-Kimmel2014,Fattahi2016a}), or to focus on the
satellite population of individual halos with masses comparable to
that of the Milky Way or M31 (see; e.g., \cite[Brook et al. 2014, Brooks et al. 2014, Wang et
al. 2015, Wetzel et al. 2016, Grand et al. 2017]{Brook2014,Brooks2014,Wang2015,Wetzel2016,Grand2017}). These simulations are able to
resolve the formation of satellite and field dwarfs down to roughly
$10^5\, M_\odot$, enabling direct comparison with the Local Group
dwarf galaxy population. I will discuss below how these simulations
resolve the ``missing satellites'' and ``too-big-to-fail'' problems. I
will focus my discussion on the results of the APOSTLE collaboration
(see; e.g., \cite[Sawala et al. 2016]{Sawala2016}, and references therein),
but similar results have also been reported by other groups. An
excellent recent review of these topics may be found in \cite[Bullock
\& Boylan-Kolchin (2017)]{Bullock2017}.

\begin{figure}[t]
\begin{center}
 \includegraphics[width=2.6in]{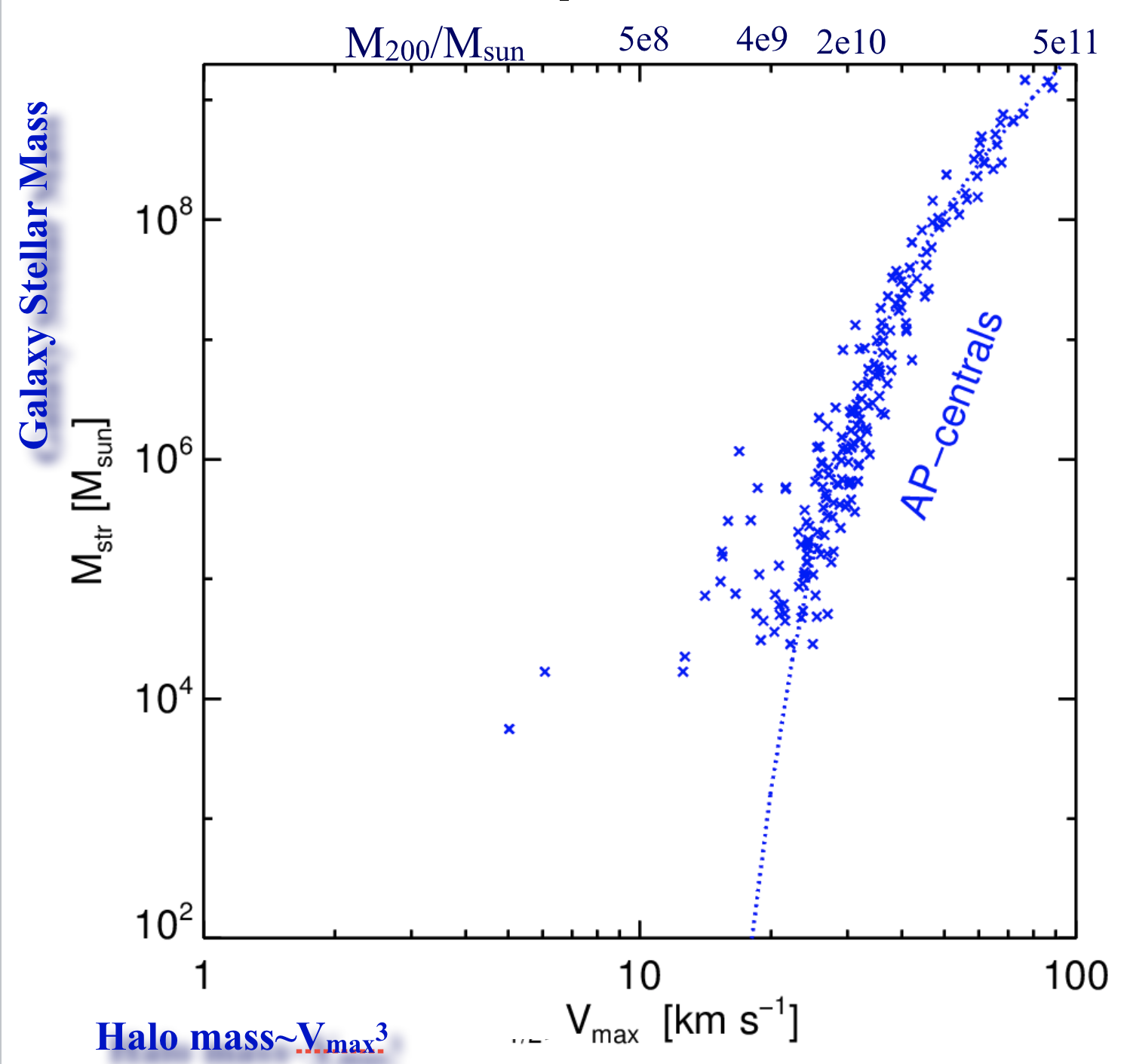} 
 \includegraphics[width=2.4in]{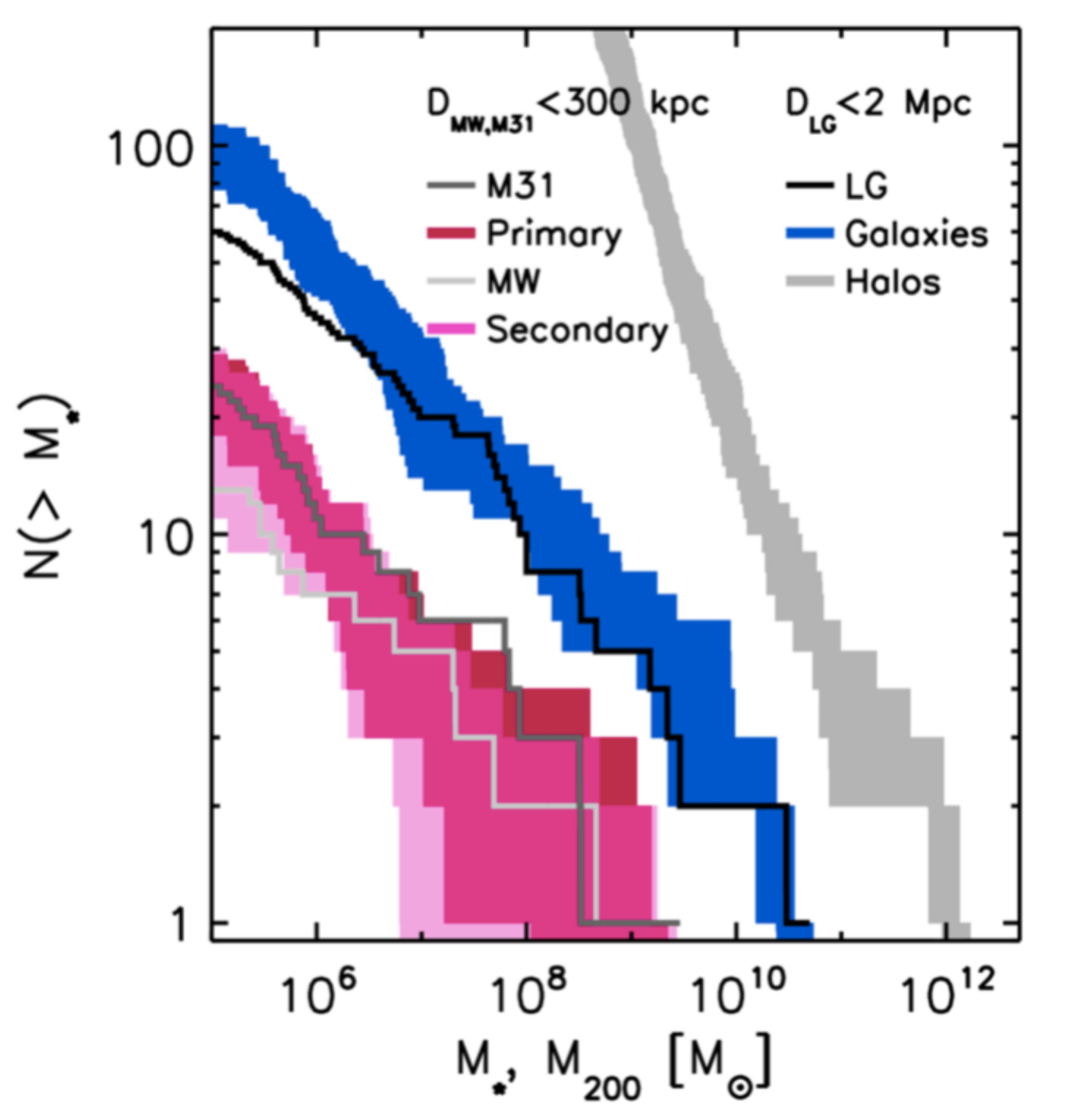} 
 \caption{Left: Stellar mass of APOSTLE field dwarfs vs halo maximum
   circular velocity ($V_{\rm max}$, scale at bottom) or virial mass ($M_{200}$, scale
   on top). Adapted from \cite[Fattahi et al. (2018)]{Fattahi2018}. Right: Cumulative number
 of APOSTLE halos and subhalos within $2$ Mpc from the Local Group barycenter
 (grey band) as a function of mass. Blue band correspond to
 luminous satellites in the same volume, as a function of stellar
 mass. Red/pink bands are satellites of the MW and M31 analogs. Solid
 lines are observational data for the Local Group. Adapted from
 \cite[Sawala et al. (2016)]{Sawala2016}.}
   \label{FigMissSats}
\end{center}
\end{figure}

\section{The Missing Satellites Problem}
\label{SecMissSats}

The left panel of Fig.~\ref{FigMissSats} shows the galaxy stellar
mass-halo mass relation for APOSTLE {\it field} dwarfs; i.e., those
that are outside the virial boundaries of any other more massive
system. Note the sharp cutoff in the relation, which implies that most
``luminous dwarfs'' (defined as those in the stellar mass range
$10^5$-$10^9\, M_\odot$) form in halos above a threshold of about
$M_{200}\sim 5\times 10^9 M_\odot$, corresponding to $V_{\rm max}\sim 20$
km/s. (The few outliers scattering towards lower masses are just
misidentified satellites, not field galaxies.) The threshold mass
arises mainly because of the effects of cosmic reionization, which
prevents gas from cooling and condensing at the center of halos whose
potential wells is shallower than that characteristic mass (see; e.g.,
\cite[Ferrero et al. 2012, Fitts et al. 2018]{Ferrero2012,Fitts2018}).

The presence of this threshold implies that, to first order, the total
number of dwarfs in the Local Group within, say, $2$ Mpc from the
MW-M31 barycenter, should be comparable to the total number of halos
and subhalos above the threshold mass. This is shown by the grey band
in the right-hand panel of Fig.~\ref{FigMissSats}, which shows the
cumulative number of halos and subhalos in APOSTLE in that volume, as
a function of mass. This indicates that the LG APOSTLE realizations
contain, on average, about $\sim 100$ halos above the threshold, and,
consequently, about $100$ dwarfs above $10^5\, M_\odot$ in stars (blue
band in same panel). By comparison, there are $60$-$70$ known dwarfs
within the same volume in the Local Group. This factor-of-two
agreement is encouraging, since such constraint was never used when
selecting the Local Group volumes for the APOSTLE project.  The same
``threshold'' mass yields between $15$ and $30$ luminous satellites
around each of the M31/MW analogs in the APOSTLE volumes (red and pink
bands), again in reasonable agreement with observation (shown with
grey lines). Note that this agreement is sensitive to the virial mass
asssumed for the LG primary galaxies. In APOSTLE, the combined virial
mass of the M31+MW system is between $2$ and
$3\times 10^{12}\, M_\odot$ (\cite[Fattahi et al
2016a]{Fattahi2016a}). For example, leaving all else unchanged, doubling
the virial mass of the primaries would double the number of
satellites, degrading the agreement between observation and the
APOSTLE results.

Note that if this mass threshold did not exist, and, for example,
luminous dwarfs could form in halos with virial masses as low mass as
$10^8 M_\odot$, then we would expect about $1,000$ such dwarfs in the
Local Group, vastly exceeding the number of known systems. This is a
clear manifestation of the ``missing satellites'' problem.
Fig.~\ref{FigMissSats} thus shows that, largely because of the effects of
reionization and feedback from evolving stars (another crucial heating
mechanism included in the simulations), the number of predicted
satellites is in good agreement with observations.  The presence of a threshold halo mass for
luminous dwarf formation thus provides a simple and compelling
resolution to the ``missing satellites'' problem in LCDM. A more thorough
discussion of this result may be found in \cite[Sawala et
al. (2016)]{Sawala2016}.

\section{The Too-Big-To-Fail Problem}
\label{SecTBTF}

The discussion above shows that there is no obvious overabundance of field
luminous dwarfs or satellites in LCDM realizations of the Local Group,
provided that dwarf galaxies form only in relatively massive
halos. This assumption may be probed observationally by estimating the
dark matter content of {\it individual} systems. In the case of dwarf
spheroidals, for example, this is accomplished by using the
line-of-sight velocity dispersion of the stars to estimate the total
mass within the stellar half-mass radius, $r_{1/2}$ (\cite[Walker et
al 2009, Wolf et al. 2010]{Walker2009,Wolf2010}). Since these are
dark matter-dominated systems, this is a direct measure of the
circular velocity of the dark matter halo at $r_{1/2}$. As discussed
in Sec.~\ref{SecIntro}, this single measurement may be used to
estimate the total mass (or maximum circular velocity, $V_{\rm max}$) of individual halos.

\begin{figure}[t]
\begin{center}
 \includegraphics[width=2.6in]{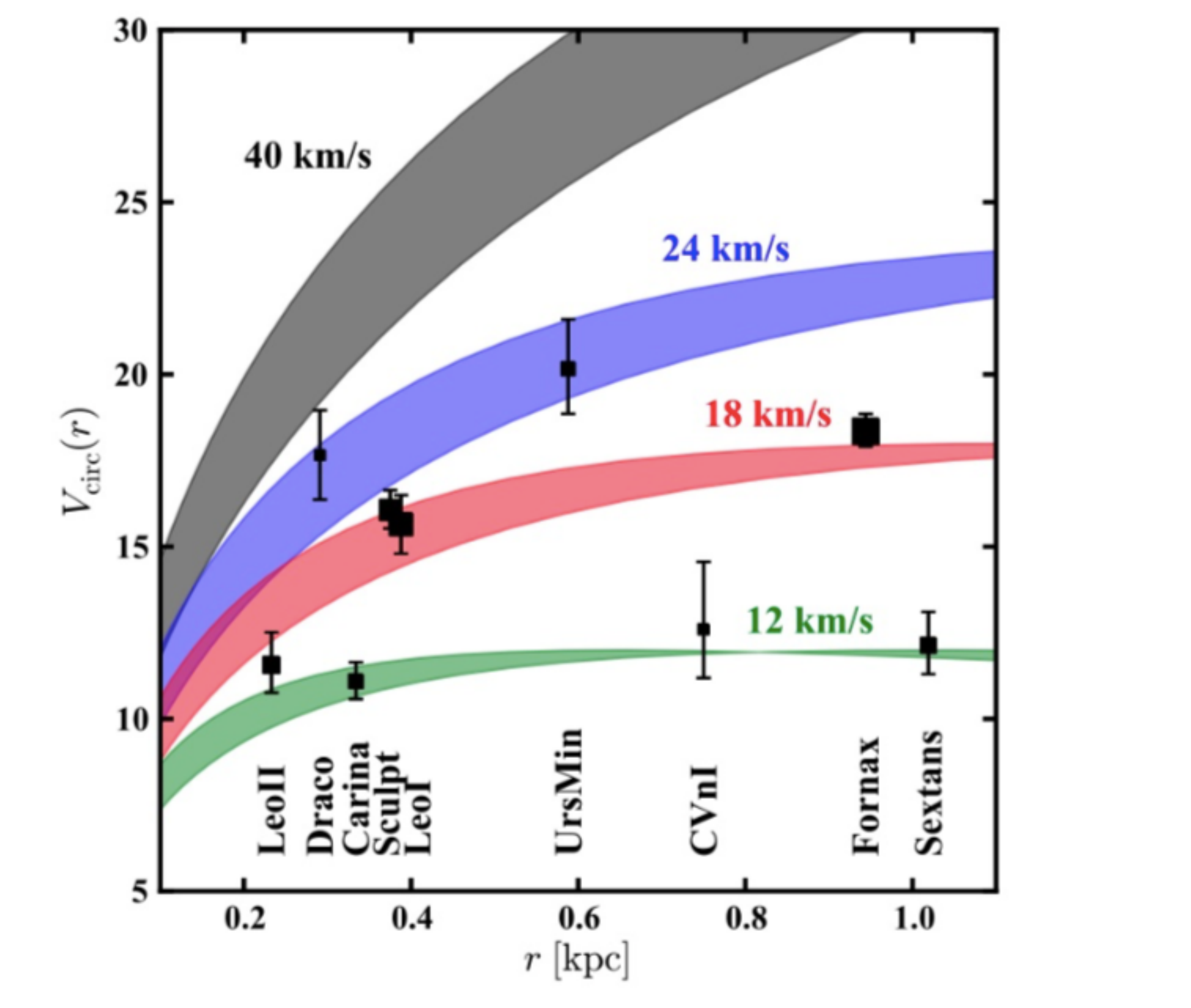} 
 \includegraphics[width=2.6in]{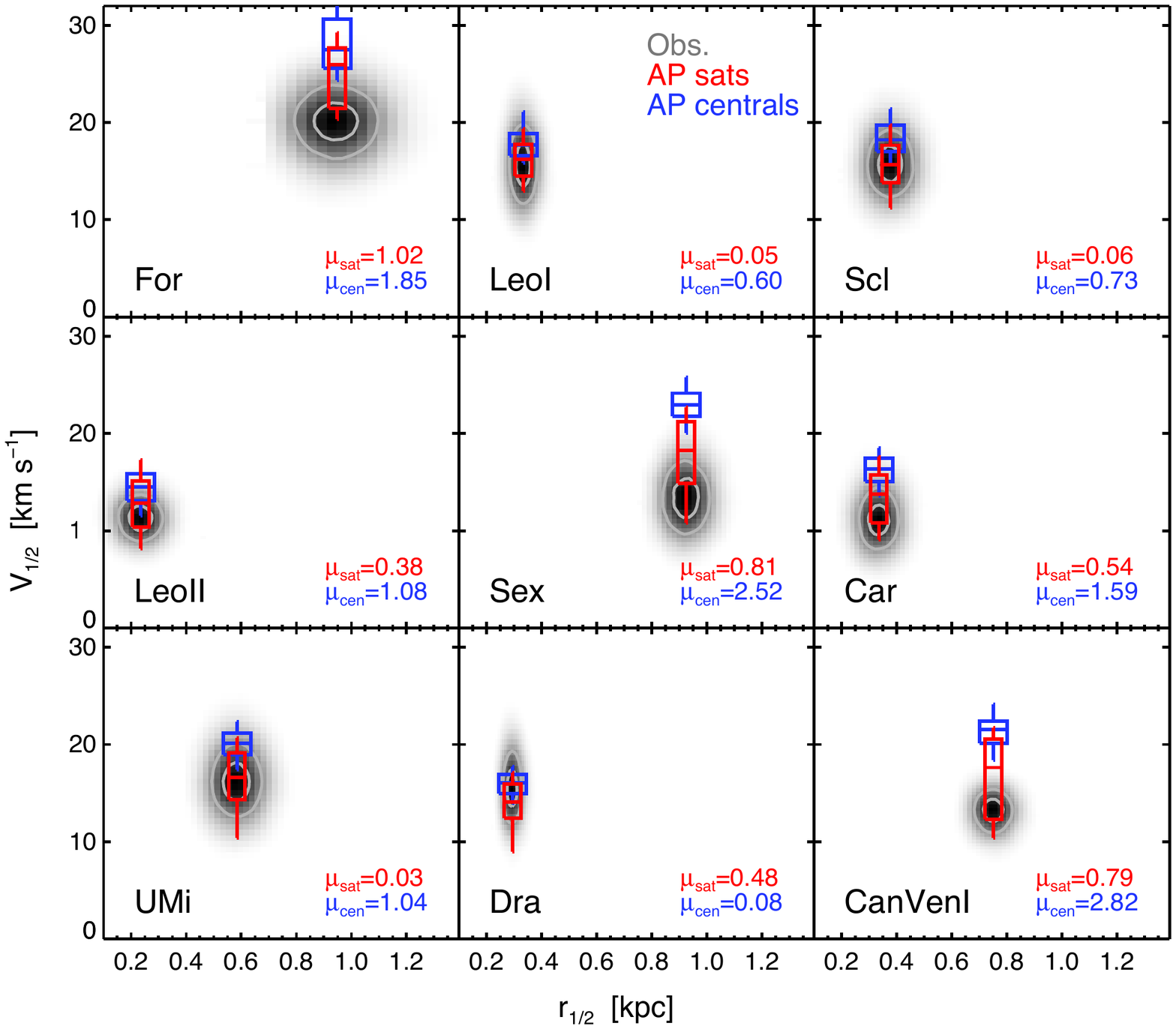} 
 \caption{Left: The estimated circular velocities at the stellar
   half-mass radii of some MW satellites, compared with the circular
   velocity curves of CDM subhalos of fixed $V_{\rm max}$, as
   specified in the legend. Adapted from
   \cite[Boylan-Kolchin et al. (2012)]{BoylanKolchin2012}. Right:  Circular velocities at
   the stellar half-mass radii of MW satellites. Grey contours
   indicate observational constraints; the red box-and-whisker symbols
 indicate the results for APOSTLE satellites of matching stellar mass at
 the same radius. Adapted from \cite[Fattahi et al. (2016b)]{Fattahi2016b}.}
   \label{FigTBTF}
\end{center}
\end{figure}

This is shown in the left panel of Fig.~\ref{FigTBTF}, where circular
velocity estimates at $r_{1/2}$ are shown for $9$ satellites of the
Milky Way (symbols with error bars). These estimates are compared with
the average circular velocity profiles of subhalos selected from the
Aquarius Project simulation suite of dark matter halo formation
(\cite[Boylan-Kolchin et al. 2012]{Boylan-Kolchin2012}). Subhalos are
binned by their maximum circular velocity, as listed in the figure legends.

This comparison illustrates a number of
important points: (i) several MW satellites appear to inhabit halos
with $V_{\rm max}<20$ km/s; i.e., below the ``threshold'' value we
discussed in Sec.~\ref{SecMissSats}; (ii) there does not seem to be a
monotonic dependence of galaxy mass with $V_{\rm max}$; for example, Draco
apparently inhabits a halo more massive than Fornax's despite being
$\sim 100\times$ less luminous; and (iii) there are a number of
subhalos that are apparently quite massive (i.e., $V_{\rm max}>25$
km/s) but yet do not have luminous counterparts. The last issue, in
particular, is what prompted \cite[Boylan-Kolchin et al (2011)]{Boylan-Kolchin2011} to argue that
such subhalos would be ``too big to fail''  (TBTF) to form luminous
satellites. This issue has often been cited as a severe observational
challenge to LCDM.

How are these puzzles explained in LCDM? Let us first address the
issue of numbers  of massive subhalos. As discussed in Sec.~\ref{SecIntro},
the number of subhalos of given mass (or $V_{\rm max}$) is a strong
function of the host halo mass. According to \cite[Wang et al
(2012)]{Wang2012}, the average number of subhalos with $V_{\rm max}$
exceeding a certain value $\nu=V_{\rm max}/V_{200}$ (where $V_{200}$
is the virial velocity of the host) is given by
$\langle N_{\rm sub} \rangle = 10.2\, (\nu/0.15)^{-3.11}$.  This
implies that we would expect roughly $10$ subhalos with
$V_{\rm max}>30$ km/s if $V_{200}\sim 200$ km/s, but only $\sim 4$ if
$V_{200}=150$ km/s.

Since (i) the virial velocity of the Milky Way is poorly constrained,
(ii) the small expected number of massive subhalos is subject to
Poisson fluctuations, and (iii) the Milky Way has at least $3$
satellites with $V_{\rm max}>30$ km/s (the LMC, SMC, and the
Sagittarius dSph), the apparent overabundance of massive subhalos that
have ``failed'' to form luminous dwarfs is intriguing but not
necessarily damning, and certainly highly sensitive to the assumed
virial mass of the Milky Way. In addition, the
$\langle N_{\rm sub} \rangle(\nu)$ expression was derived for dark
matter only simulations, and should be corrected when applied to
dwarfs. The correction arises because dwarf galaxy halos expel most of
their baryons at early times, reducing the depth of their potential
wells, and reducing the values of $\nu$ by $\sim 20\%$ (\cite[Sawala
et al 2016]{Sawala2016}). Because of the steepness of the subhalo mass
function (see, e.g., the right-hand panel of Fig.~\ref{FigMissSats}),
even this modest correction leads to a factor of about two reduction
in the expected number of massive subhalos. Considering all of these
arguments, it is hard to conclude that the number of subhalos more
massive than $V_{\rm max}\sim 30$ km/s poses a substantial challenge
to LCDM.

The above discussion addresses only one of the three items that, in my
opinion, define the TBTF conundrum. The other two issues (i.e., why
satellites populate halos below the ``threshold'' mass, and
why the dependence between halo mass and stellar
mass is not monotonic) are explained by the effects of tidal
stripping by the host halo, as well as by the very steep dependence of
stellar mass on halo mass shown in the left-hand panel of
Fig.~\ref{FigMissSats}. This panel shows that it is indeed possible for dwarfs
of widely differing stellar mass to be formed in
halos of similar virial mass. If affected by tides differently, then a more
luminous satellite could easily today have less dark matter than a fainter one,
as is the case of Fornax and Draco.

This is because tides tend to strip less bound material first, leading to a
reduction in the dark matter halo content of a satellite. The
reduction affects not only the outskirts of a satelite but also
the dark matter {\it within} the stellar half-mass radius, even when
the stellar component remains bound. This happens because, unlike stars,
dark matter particles found {\it within} $r_{1/2}$ have apocentric radii far
outside the luminous radius of the satellite and are therefore
more vulnerable to stripping. In loose language, these inner particles get stripped
while they are at apocenter, and are not replaced, leading to a
reduction of dark matter inside $r_{1/2}$ (see; e.g., \cite[Penarrubia
et al 2008]{Penarrubia2008}).

Are tidal effects enough to reconcile the data in the left panel of
Fig.~\ref{FigTBTF} with LCDM? This has been examined independently by
a number of authors who broadly agree that they do, although there is
still some debate about the exact role of halo mass, small number
statistics, baryon-induced core formation, and tidal stripping (see,
e.g., the discussion and references in \cite[Fattahi et al
2016b]{Fattahi2016b}, which complements and extends that of \cite[Sawala
et al 2016]{Sawala2016}). The analysis in many of those papers is
statistical, and largely based on reproducing the left-hand panel of
Fig.~\ref{FigTBTF} with data from hydrodynamical simulations to argue
that there are no large numbers of obvious ``failures'' (i.e., curves
that are always above the observational data). 

The recent analysis of \cite[Fattahi et al (2016b)]{Fattahi2016b} goes beyond that, and
compares the predictions of APOSTLE satellite and field dwarfs with
each of the nine MW satellites. This is shown in the right-hand panel
of Fig.~\ref{FigTBTF}, where the grey contoured area indicate the
observed circular velocity constraint for each individual
satellite, as well as the predictions from the APOSTLE simulations
{\it at the same radius} for systems matching the stellar mass of each
satellite. Blue box-and-whisker symbols are predictions for field
dwarfs; red ones indicate results for satellite systems. Note that red
symbols are always below the blue ones, as expected for tidal
stripping of dark matter. Overall, there is fair agreement between the
APOSTLE results and observations, for all individual MW satellites. We
conclude that, when including tidal effects, there is no obvious
difficulty matching the observed constraints with the predictions of
LCDM cosmological hydrodynamical simulations. The ``too-big-to-fail''
LCDM problem has thus, in my opinion, been successfuly resolved.

\section{RELHICs: Reionization-Limited HI Clouds}

One important consequence of the threshold virial mass for luminous
dwarf galaxy formation discussed in the previous subsections is that there
should exist some halos just below the threshold  (i.e., ``mini-halos'' with
$M_{200} \sim 3\times 10^8$ to $\sim 5\times 10^9 \, M_\odot$) which, although
unable to form stars, are still able to retain and bind some of the
gas heated by cosmic reionization. Indeed, it is straightforward to
compute the total gas mass, density profile, and temperature profile
of photoionized primordial gas bound, at $z=0$, to a low-mass LCDM
halo. This gas (i)
should be near hydrostatic equilibrium with the gravitational
potential of the dark matter; (ii) should be essentially free of metals;
(iii) have negligible velocity dispersion; and (iii) be confined by the gravity of
the halo and the ambient pressure of the ionized intergalactic medium
(\cite[Benitez-Llambay et al 2017]{Benitez-Llambay2017}).

\begin{figure}[t]
\begin{center}
 \includegraphics[width=2.5in]{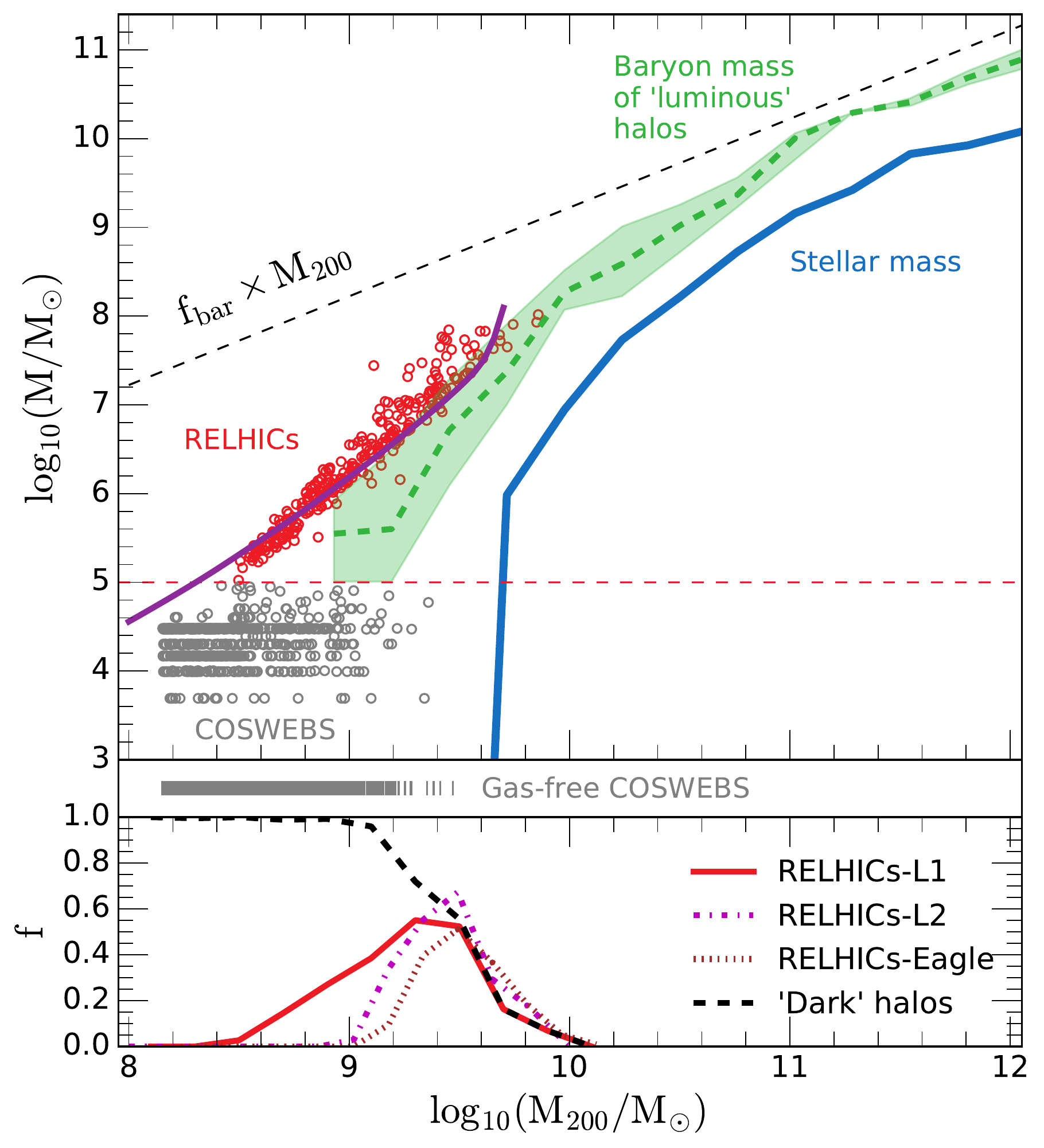} 
 \includegraphics[width=2.7in]{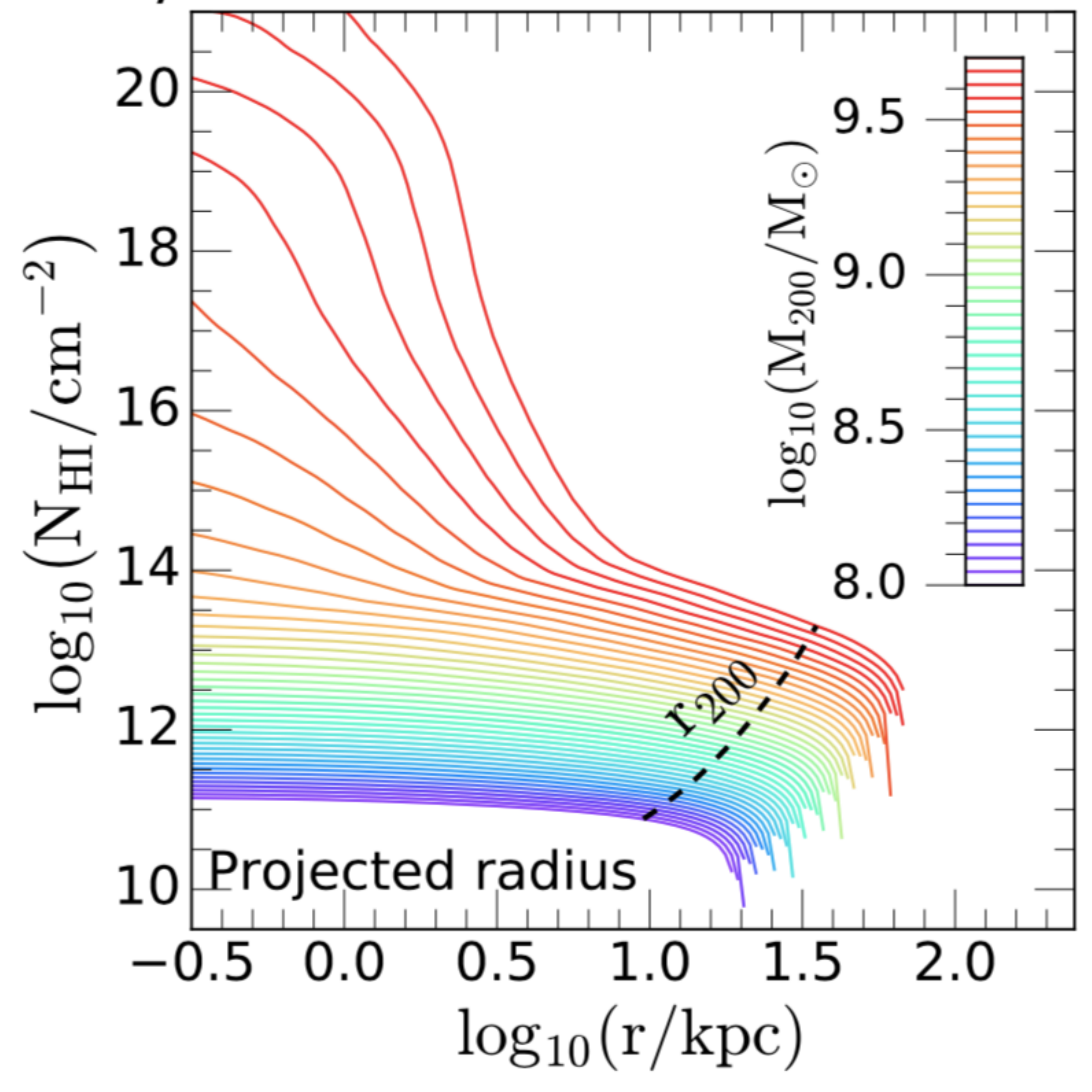} 
 \caption{ Left: The virial mass dependence of the baryonic mass of
   galaxies in the APOSTLE simulation suite. The blue solid line
   traces the median galaxy stellar mass. Note the threshold at
   $\sim 5\times 10^9\, M_\odot$, below which no luminous galaxies
   form. The red and grey open circles indicate the baryonic (gas) mass
   within the virial radius of ``dark'' halos. Right: The HI column
   density profile of RELHICs in the APOSTLE simulations. The color
   bar indicates the virial mass of the surrounding halo.  Adapted
   from \cite[Benitez-Llambay et al (2017)]{Benitez-Llambay2017}. }
   \label{FigRELHICs}
\end{center}
\end{figure}

The total gas mass expected within the virial radius of ``mini-halos''
is shown, as a function of halo virial mass, by the magenta solid line
in the top-left panel of Fig.~\ref{FigRELHICs}. The total gas mass bound
to such low-mass halos is, as expected, well below the universal
cosmic average, which is indicated by the dashed line. The virial mass
upper limit of these ``RELHICs'' (REionization-Limited HI Clouds) is
determined by the central cooling time of the gas, which becomes
shorter than the age of the Universe above
$\sim 5\times 10^9 \, M_\odot$. This corresponds, in
Fig.~\ref{FigRELHICs}, to the threshold halo mass above which star
formation can proceed and luminous galaxies form (see, e.g., blue
solid line).

The APOSTLE simulations contain a large number of low-mass halos in
the ``mini-halo'' mass range. Some of these halos contain RELHICs;
(red open circles in the left panel of Fig.~\ref{FigRELHICs}), but
others are basically gas-free\footnote{Recall that one gas particle in the
highest-resolution APOSTLE runs corresponds to about
$10^4\, M_\odot$.}  (COSWEBS, or ``cosmic web-stripped
systems''; grey open circles). As discussed in \cite[Benitez-Llambay et
al. (2017)]{Benitez-Llambay2017}, the difference between RELHICs and
COSWEBS is their location within the Local Group. Mini-halos that are
relatively close to the two LG primary galaxies have had their gas
content efficiently removed by ram-pressure from the ``cosmic web''
(\cite[Benitez-Llambay et al 2013]{Benitez-Llambay2013}), which gives origin to the COSWEBS population.

On the other hand, mini-halos in low-density environments well away
from the primaries are able to retain their gas, and constitute
excellent examples of the RELHICs population hypothesized above. Their
thermodynamic properties are well specified, and their gas density and
temperature profiles may be predicted in detail. Gas in RELHICs is
nearly fully ionized but with neutral cores that span a large range of
H I masses and column densities and have negligible non-thermal
broadening. Their predicted HI column density profiles are shown in
the right-hand panel of Fig.~\ref{FigRELHICs}.

A full analysis of the simulated RELHICs population in APOSTLE is
provided in \cite[Benitez-Llambay et al (2017)]{Benitez-Llambay2017}\footnote{Note that the ALFALFA
  data in Fig.~10 of that paper is incorrectly plotted and that its
  discussion will soon be revised in an erratum. My thanks to Yakov
  Faerman and Betsy Adams for pointing this out.}, who argue that
Local Group RELHICs (i) should typically be beyond 500 kpc from the
Milky Way or M31; (ii) have positive Galactocentric radial velocities;
(iii) HI sizes not exceeding 1 kpc, and (iv) should be nearly
round. Indeed, it is possible that some have already been detected in
blind HI surveys, like ALFALFA, which have identified Ultra Compact
High Velocity HI Clouds (\cite[Adams et al. 2013]{Adams2013}). The simulations provide
guidance to identify which of those systems might be ``mini-halos''
and which are just random condensations of neutral gas in the Galactic
halo. One way of discriminating between the two would be to measure
H$_\alpha$ emission in some RELHICs candidates, which should show a
distinct ``ring'' marking the sharp transition between the inner
neutral core and the outer ionized envelop of RELHICs. No such object
has been reported yet, but it should be clear that the detection and
characterization of RELHICs would offer a unique and exciting probe of
the small-scale clustering of cold dark matter.

\section{Concluding Remarks} 

The discussion above shows how the properties of dwarf galaxies may be
used to probe some of the distinctive predictions of the LCDM
cosmological paradigm on small scales. I have reviewed how
cosmological hydrodynamical simulations of dwarf galaxy formation have
helped to clarify the interpretation of some observations that are often
cited as major ``challenges'' to LCDM. In
particular, I showed how the ``missing satellites'' and
``too-big-to-fail'' problems can be resolved without appealing to {\it
  any} modification to the cold dark matter paradigm.

This short review is, like any and all, incomplete, and does not address the full
list of small-scale LCDM worries that may be found in the
literature. For example, I have not discussed the ``cusp vs core''
controversy but recent work has highlighted promising progress on that
respect (see, e.g., the contribution of Alyson Brooks elsewhere in
this volume). Less well understood are potential challenges that may
arise from some of the faintest satellites of the Milky Way (i.e., the
``ultra-faint'' population of galaxies with stellar masses well below
$10^5\, M_\odot$), which are still beyond the capabilities of current
simulations. According to the discussion above, these ultra-faints
should also form in fairly massive halos, at odds with the very low
velocity dispersion measured for some. This may, in principle, be
resolved by appealing to the effects of extreme tidal stripping (see;
e.g., the discussion in \cite[Fattahi et al 2018]{Fattahi2018}), but
the issue is far from settled so far.

Another issue that is far from settled concerns the origin of the
morphological diversity, scalings, and scatter in the dwarf galaxy
population. What sets their size and stellar mass profiles? Why do
some rotate and others do not? Why do some have gas and others not?
What determines their star formation history? Is the observational
evidence consistent with the sharp threshold in virial mass for dwarf
formation espoused above? These are issues that concern us now and
will keep us busy for some time to come. Making sense of the wondrous
diversity of dwarf galaxies, and contrasting it with the relatively
featureless and self-similar context of their dark halos, is bound to
yield interesting clues not only about the nature of dark matter, but
also of the physics behind the assembly of the faintest galaxies.


\begin{thebibliography}{}


\bibitem[Adams et al.(2013)]{2013ApJ...768...77A} Adams, E.~A.~K., Giovanelli, R., \& Haynes, M.~P.\ 2013, \textit{ApJ}, 768, 77 

\bibitem[Behroozi et al.(2013)]{Behroozi2013} Behroozi, P.~S.,
  Marchesini, D., Wechsler, R.~H., et al.\ 2013, \textit{ApJL}, 777,
  L10 

\bibitem[Ben{\'{\i}}tez-Llambay et al.(2013)]{2013ApJ...763L..41B} Ben{\'{\i}}tez-Llambay, A., Navarro, J.~F., Abadi, M.~G., et al.\ 2013, \textit{ApJL}, 763, L41 


\bibitem[Ben{\'{\i}}tez-Llambay et al.(2017)]{2017MNRAS.465.3913B} Ben{\'{\i}}tez-Llambay, A., Navarro, J.~F., Frenk, C.~S., et al.\ 2017, \textit{MNRAS}, 465, 3913 

\bibitem[Boylan-Kolchin et al.(2011)]{2011MNRAS.415L..40B} Boylan-Kolchin, M., Bullock, J.~S., \& Kaplinghat, M.\ 2011, \textit{MNRAS}, 415, L40 

\bibitem[Boylan-Kolchin et al.(2012)]{2012MNRAS.422.1203B} Boylan-Kolchin, M., Bullock, J.~S., \& Kaplinghat, M.\ 2012, \textit{MNRAS}, 422, 1203 


\bibitem[Brook et al.(2014)]{Brook2014} Brook, C.~B., Di Cintio, A., Knebe, A., et al.\ 2014, \textit{ApJL}, 784, L14 


\bibitem[Brooks \& Zolotov(2014)]{2014ApJ...786...87B} Brooks, A.~M., \& Zolotov, A.\ 2014, \textit{ApJ}, 786, 87 


\bibitem[Bullock \& Boylan-Kolchin(2017)]{2017ARA&A..55..343B} Bullock, J.~S., \& Boylan-Kolchin, M.\ 2017, \textit{ARAA}, 55, 343 


\bibitem[Fattahi et al.(2016a)]{Fattahi2016a} Fattahi, A., Navarro, J.~F., Sawala, T., et al.\ 2016a, \textit{MNRAS}, 457, 844 


\bibitem[Fattahi et al.(2016b)]{Fattahi2016b} Fattahi, A., Navarro, J.~F., Sawala, T., et al.\ 2016b, arXiv:1607.06479 


\bibitem[Fattahi et al.(2018)]{2018MNRAS.476.3816F} Fattahi, A., Navarro, J.~F., Frenk, C.~S., et al.\ 2018, \textit{MNRAS}, 476, 3816 



\bibitem[Ferrero et al.(2012)]{2012MNRAS.425.2817F} Ferrero, I., Abadi, M.~G., Navarro, J.~F., Sales, L.~V., \& Gurovich, S.\ 2012, \textit{MNRAS}, 425, 2817 


\bibitem[Fitts et al.(2017)]{2017MNRAS.471.3547F} Fitts, A.,
  Boylan-Kolchin, M., Elbert, O.~D., et al.\ 2017, \textit{MNRAS},
  471, 3547 


\bibitem[Frenk \& White(2012)]{2012AnP...524..507F} Frenk, C.~S., \& White, S.~D.~M.\ 2012, \textit{Annalen der Physik}, 524, 507 

\bibitem[Garrison-Kimmel et al.(2014)]{Garrison-Kimmel2014} Garrison-Kimmel, S., Boylan-Kolchin, M., Bullock, J.~S., \& Lee, K.\ 2014, \textit{MNRAS}, 438, 2578 


\bibitem[Grand et al.(2017)]{2017MNRAS.467..179G} Grand, R.~J.~J., G{\'o}mez, F.~A., Marinacci, F., et al.\ 2017, \textit{MNRAS}, 467, 179 



\bibitem[McConnachie(2012)]{McConnachie2012} McConnachie, A.~W.\ 2012, \textit{AJ}, 144, 4 



\bibitem[Moore et al.(1999)]{Moore1999} Moore, B., Ghigna, S., Governato, F., et al.\ 1999, \textit{ApJL}, 524, L19 

\bibitem[(Navarro et al. 1996)]{NFW96} Navarro, J.~F., Frenk, C.~S., \& White, S.~D.~M.\ 1996, \textit{ApJ}, 462, 563 

\bibitem[(Navarro et al. 1997)]{NFW97} Navarro, J.~F., Frenk, C.~S., \& White, S.~D.~M.\ 1997, \textit{ApJ}, 490, 493 


\bibitem[Pe{\~n}arrubia et al.(2008)]{2008ApJ...673..226P} Pe{\~n}arrubia, J., Navarro, J.~F., \& McConnachie, A.~W.\ 2008, \textit{ApJ}, 673, 226 



\bibitem[Planck Collaboration et al.(2016)]{Planck2016} Planck Collaboration, Ade, P.~A.~R., Aghanim, N., et al.\ 2016, \textit{A\&A}, 594, A13 


\bibitem[Sawala et al.(2016)]{2016MNRAS.457.1931S} Sawala, T., Frenk, C.~S., Fattahi, A., et al.\ 2016, \textit{MNRAS}, 457, 1931 

\bibitem[Schaye et al.(2015)]{Schaye2015} Schaye, J., Crain, R.~A., Bower, R.~G., et al.\ 2015, \textit{MNRAS}, 446, 521 

\bibitem[Vogelsberger et al.(2014)]{Vogelsberger2014} Vogelsberger, M., Genel, S., Springel, V., et al.\ 2014, \textit{Nature}, 509, 177 


\bibitem[Walker et al.(2009)]{2009ApJ...704.1274W} Walker, M.~G., Mateo, M., Olszewski, E.~W., et al.\ 2009, \textit{ApJ}, 704, 1274 




\bibitem[Wang et al.(2012)]{Wang2012} Wang, J., Frenk, C.~S., Navarro, J.~F., Gao, L., \& Sawala, T.\ 2012, \textit{MNRAS}, 424, 2715 


\bibitem[Wang et al.(2015)]{2015MNRAS.454...83W} Wang, L., Dutton, A.~A., Stinson, G.~S., et al.\ 2015, \textit{MNRAS}, 454, 83 


\bibitem[Wolf et al.(2010)]{2010MNRAS.406.1220W} Wolf, J., Martinez, G.~D., Bullock, J.~S., et al.\ 2010, \textit{MNRAS}, 406, 1220 



\end{thebibliography}
\end{document}